\documentclass[conference]{IEEEtran}
\IEEEoverridecommandlockouts
\usepackage{tikz}
\usetikzlibrary{positioning}
\usepackage[T1]{fontenc}
\usepackage{cite}
\usepackage{amsmath,amssymb,amsfonts}
\usepackage{algorithmic}
\usepackage{graphicx}
\usepackage{textcomp}
\usepackage{xcolor}
 \usepackage{caption}
\usepackage{subcaption}
\usepackage{cleveref}
\usepackage{booktabs}
\usepackage{float}
\def\BibTeX{{\rm B\kern-.05em{\sc i\kern-.025em b}\kern-.08em
    T\kern-.1667em\lower.7ex\hbox{E}\kern-.125emX}}

\def\layersep{3cm}
\newcommand\nn[1]{
    \foreach \y in {1,2,3,5}
        \node[neuron, fill=green!40] (X\y-#1) at (0 ,\y+1) {$X\y$};

    \foreach \y in {1,...,5}
        \path node[neuron, fill=blue!40] (H\y-#1) at (\layersep,\y) {$H\y$};

    \node[neuron, fill=red!40] (Y-#1) at (2*\layersep,2.5) {$Y$};

    \foreach \source in {1,2,3,5}
        \foreach \dest in {1,...,5}
            \path (X\source-#1) edge (H\dest-#1);

    \foreach \source in {1,...,5}
        \path (H\source-#1) edge (Y-#1);
}

\begin{document}

\title{Detection of Electric Motor Damage Through Analysis of Sound Signals Using Bayesian Neural Networks\\
\thanks{Work partially realised in the scope of project titled ''Process Fault Prediction and Detection''. Project was financed by The National Science Centre on the base of decision no. UMO-2021/41/B/ST7/03851. Part of work was funded by AGH’s Research University Excellence Initiative under projects: Waldemar Bauer PAKIET HABILITACYJNY -edycja II  and “DUDU - Diagnostyka Uszkodzeń i Degradacji Urządzeń''. 
Marta Zagorowska gratefully acknowledges funding from Marie Curie Horizon Postdoctoral Fellowship project RELIC, grant no 101063948 (writing and revisions).}
}

\author{
\IEEEauthorblockN{Waldemar Bauer}
\IEEEauthorblockA{\textit{Dept. of Automatic Control \& Robotics} \\ 
\textit{AGH University of Krakow}\\
Krakow, Poland \\
bauer@agh.edu.pl}
\and
\IEEEauthorblockN{Marta Zagórowska}
\IEEEauthorblockA{\textit{Dept. of Engineering Cybernetics} \\ 
\textit{Norwegian University} \\\textit{of Science and Technology}\\
Trondheim, Norway \\
marta.zagorowska@ntnu.no}
\and
\IEEEauthorblockN{Jerzy Baranowski}
\IEEEauthorblockA{\textit{Dept. of Automatic Control \& Robotics} \\ 
\textit{AGH University of Krakow}\\
Krakow, Poland \\
jb@agh.edu.pl}
}

\maketitle

\begin{abstract}
Fault monitoring and diagnostics are important to ensure reliability of electric motors. Efficient algorithms for fault detection improve reliability, yet development of cost-effective and reliable classifiers for diagnostics of equipment is challenging, in particular due to unavailability of well-balanced datasets, with signals from properly functioning equipment and those from faulty equipment. Thus, we propose to use a Bayesian neural network to detect and classify faults in electric motors, given its efficacy with imbalanced training data. The performance of the proposed network is demonstrated on real life signals, and a robustness analysis of the proposed solution is provided.\end{abstract}
\begin{IEEEkeywords}
Bayesian Neural Networks, Fault detection, signal, engine, commutator motors
\end{IEEEkeywords}

\section{Introduction}
Fault detection is an essential task in numerous industrial applications such as manufacturing, transportation, and energy production. Faults in these applications can lead to catastrophic events such as system failure, increased maintenance costs, and even loss of life. Even though fault detection has been an area of extensive research in recent years, development of cost-effective and reliable classifiers for fault detection in electric motors is still an unresolved issue. In particular, collection of a well-balanced dataset with signals from functioning and faulty equipment remains a challenge. Thus, the concept of employing a Bayesian neural network (BNN) is proposed, given its efficacy with imbalanced training data. 

Currently, the problem of detecting motor faults is solved by DNNs using different types of signals. The article \cite{Tran2023} shows how to detect faults in a servo mechanism system using acceleration measurement and DNN. The authors of \cite{Qiu2023} show a similar solution but for the current signal. It is also possible to use a vibration signal for this process, an example of which is \cite{Tama2022}. An example of an approach to fault diagnostics in a data fusion system can be seen in \cite{Zhang2023}. The work \cite{Zhu2023} presents a review of machine learning methods for the problem of fault detection in electric motors.

Bayesian neural networks (BNNs) are a type of neural network that integrates Bayesian inference to estimate the distribution of weights in the network \cite{mac2003}. By incorporating prior knowledge about the data and propagating uncertainty throughout the network, BNNs can provide more robust predictions and better account for uncertainty in the input data \cite{10.5555/3045390.3045502,Hinton2012}. In recent years, BNNs have been applied to fault detection in various industrial applications, and several studies have shown their effectiveness. For instance, in a study on wind turbine fault detection, in \cite{Bakri2017} proposed a BNN-based fault detection system that can handle sparse data and provide better uncertainty estimates. The system outperformed traditional methods and reduced false alarms significantly. The BNN-based fault detection system for gas turbine engines was described in \cite{Loboda2015}. They demonstrated that the BNN-based system outperformed traditional neural networks and provided more accurate fault detection. Similarly in \cite{Ma2020} proposed a fault detection system for chemical processes based on BNNs with dropout. The system produced more accurate fault detection results than traditional neural networks.

While existing approaches demonstrated theoretical potential of BNNs in fault detection, there was little work done on signal availability and representation for fault detection. To address this, the creation of a classifier capable of rapid fault diagnosis using acoustic data collected by commonly available devices such as smartphones is highly advantageous. Thus, the concept of employing a Bayesian neural network (BNN) based on frequency domain instead of time domain is proposed in this paper. The contributions of the paper are twofold:
\begin{itemize}
\item A method was proposed for acoustic signal representation for damage detection purposes based on frequency domain,
\item A structure of the Bayesian neural network for detecting damage was proposed and analysed,
\end{itemize}
The new framework was validated on real sound signals collected using a smartphone from domestic appliances.
 
The paper is arranged in the following manner. \Cref{sec:BNN} provide an overview of the Bayesian inference approach used throughout the paper, covering the fundamentals of the Bayesian Neural Network method, data set characteristics and the schema of used Bayesian Neural Network. Then we follow with the description of considered case study in \Cref{sec:CaseStudy} and results of BNNs on the sound signals set from real devices \Cref{sec:FaultDetection}. We finish the paper with the conclusions in \Cref{sec:Conclusions}.

\section{Bayesian neural networks}
\label{sec:BNN}

Bayesian Neural Networks (BNNs) are a type of neural network that use Bayesian inference to provide a probabilistic approach to neural network learning. In contrast to traditional neural networks, which use a point estimate of weights, BNNs learn a probability distribution over the weights. This distribution allows us to incorporate prior knowledge and uncertainty into our model \cite{Rezende2014,mac2003}.

Suppose we have a training dataset:
\[
D = \lbrace(x_1, y_1), (x_2, y_2), …, (x_n, y_n)\rbrace,
\]
where $x_i$ is the $i-$th input and $y_i$ is the corresponding output. Let $w$ denote the weights of the neural network. In a traditional neural network, we would try to find a set of weights $w$ that minimize the loss function $L(D,w)$ over the training dataset:

\begin{equation}
      L(D,W) = \sum_{x_i,y_i}(y_i - f(x_i,w))^2+ \lambda\sum_{d}w_{d}^2 
\end{equation}

In a BNN, we instead learn a probability distribution over the weights:

\begin{equation}
    \mathrm{log}\,p(D,w) = \sum_{x_i,y_i}\mathrm{log}N(y_i | f(x_i,w),I)\\ + \sum_{d}\mathrm{log}N(w_d |0,\cfrac{1}{\lambda})
\end{equation}
where $p(D,w)$ is the likelihood of the data given the weights, $p(w)$ is the prior distribution over the weights, and $p(D)$ is the marginal likelihood of the data. 

BNNs provide a way to incorporate uncertainty and prior knowledge into neural network learning by using Bayesian inference to learn a probability distribution over the weights. This distribution allows us to make predictions and calculate uncertainties that reflect the model's uncertainty about the true values of the weights. Using Bayes' theorem, we can compute the posterior distribution over the weights:
\begin{equation}
\begin{aligned}
p(w|D) =&{} \cfrac{p(D|w)p(w)}{p(D)}\\ =&{} \cfrac{p(D|w)p(w)}{\int_{w'}p(D|w')p(w')\mathrm{d}w'} 
\end{aligned}
\end{equation}
The posterior distribution gives us a distribution over possible values for the weights, which we then use to make predictions and calculate uncertainties.

To make predictions on a new input $x$, we compute the posterior predictive distribution over the output $\hat{y}$:
\begin{equation}
\label{eq:Integral}
\begin{aligned}
p(\hat{y}(x)|D) =&{} \int_{w} p(\hat{y}(x)|w)p(w|D)\mathrm{d}w\\ =&{}\mathbb{E}_{p(w|D)}[p(\hat{y}(x)|w]
\end{aligned}
\end{equation}
The integral in \eqref{eq:Integral} averages the predictions over all possible values of the weights, weighted by their posterior probabilities.

\section{Case study} 
\label{sec:CaseStudy}

\subsection{Dataset and faults}
\begin{figure}[!tb]
     \centering
          \begin{subfigure}[b]{0.42\textwidth}
         \centering
  \includegraphics[width=1\columnwidth]{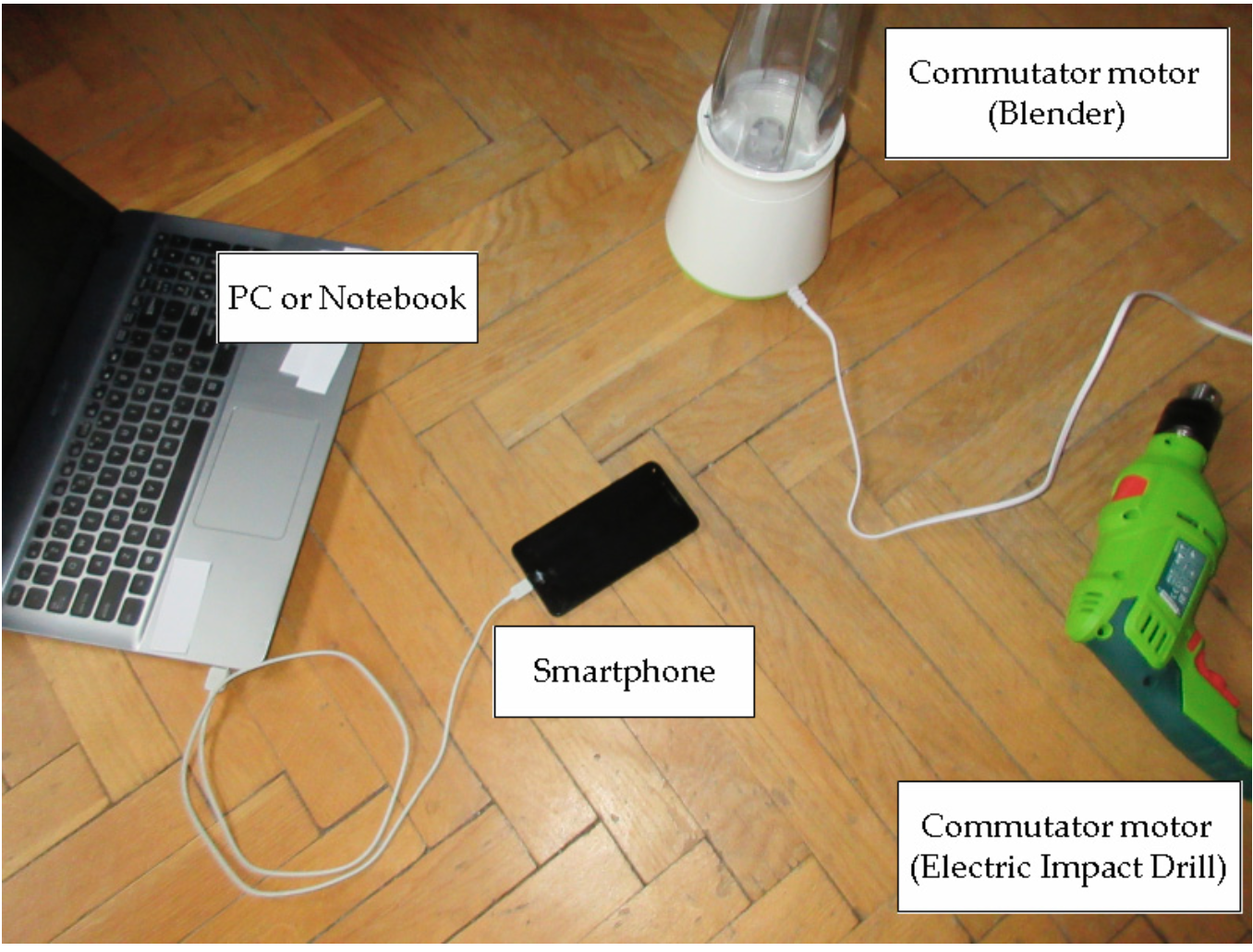}
\caption{Experimental setup\label{fig:aqui}}
     \end{subfigure}
     \begin{subfigure}[b]{0.42\textwidth}
     \centering
  \includegraphics[width=1\columnwidth]{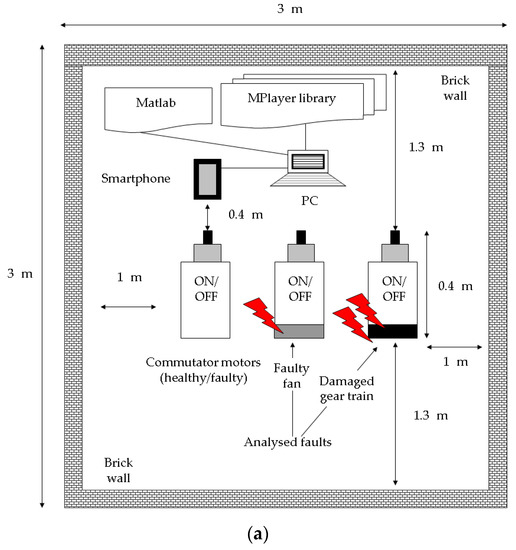}
\caption{Data acquisition setup\label{fig:aqui_1}}
     \end{subfigure}
     \hfill
        \caption{Acoustic data acquisition from blender and electric impact drill using a smartphone (CC BY 4.0, source \cite{glowacz2018})}
        \label{fig:DataAcquisition}
\end{figure}

\begin{figure*}[!tb]
     \centering
     \begin{subfigure}[!tb]{0.42\textwidth}
         \centering
  \includegraphics[width=1\columnwidth]{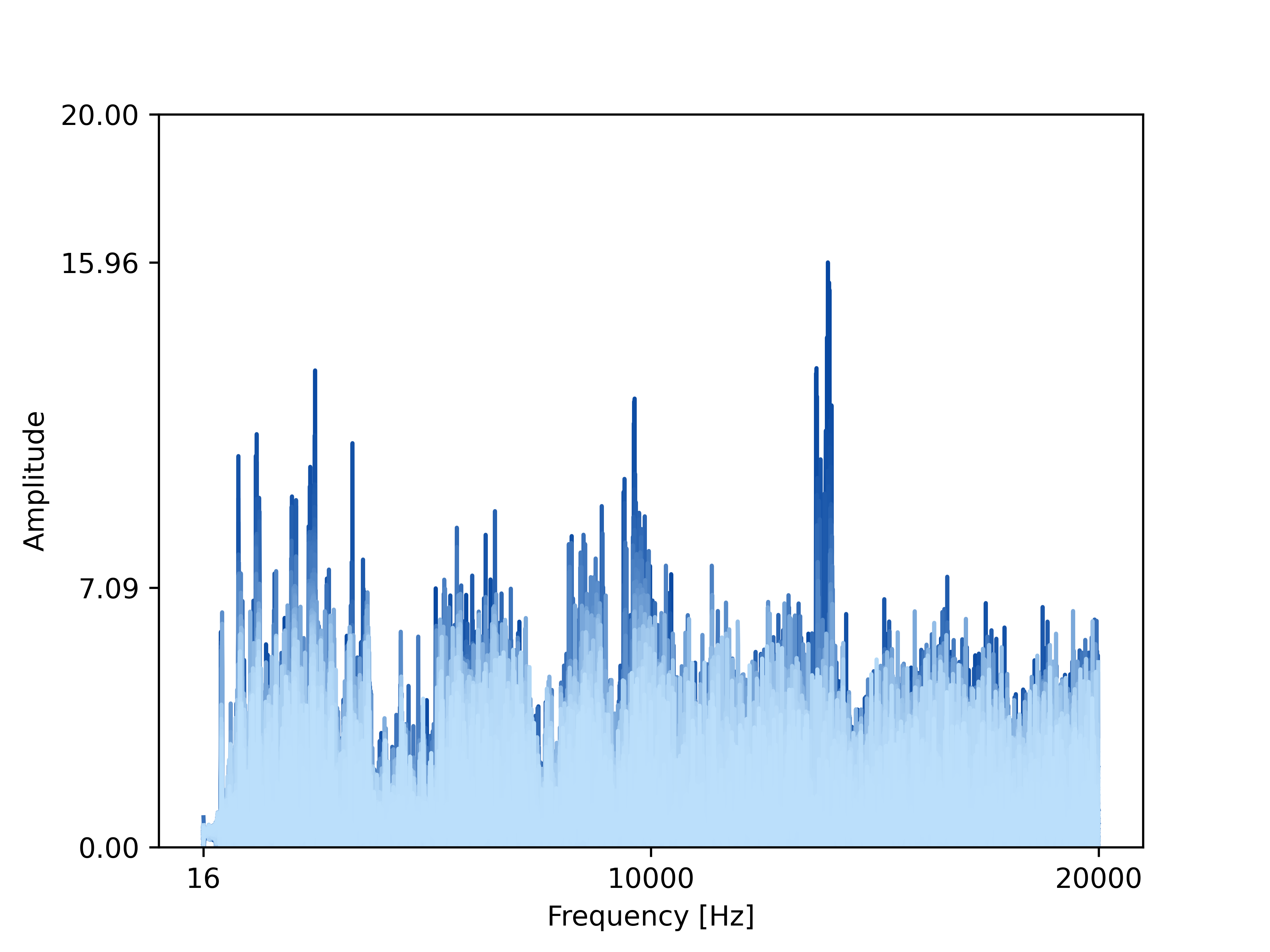}
\caption{FFT transforms for signals of properly functioning devices.\label{fig:hsig}}
     \end{subfigure}
     \hfill
     \begin{subfigure}[!tb]{0.42\textwidth}
         \centering
  \includegraphics[width=1\columnwidth]{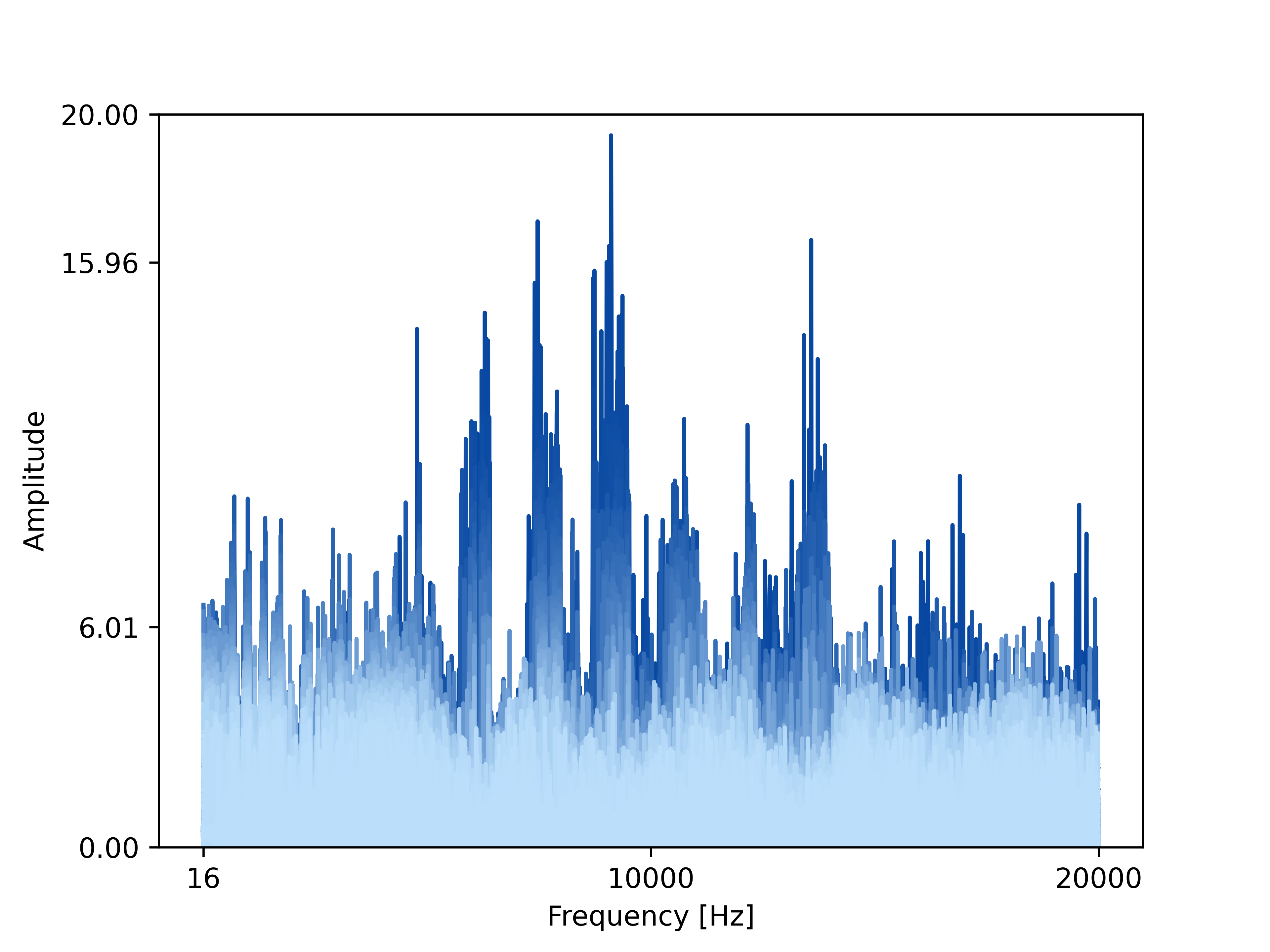}
\caption{FFT transforms for signals of faulty devices. \label{fig:uhsig}}
     \end{subfigure}
     \hfill
        \caption{FFT transforms. The light blue color of the plot represents signals with small maximum amplitudes, while the dark blue color corresponds to signals with large maximum amplitudes}
        \label{fig:FFT}
\end{figure*}

The experiments will utilize a dataset obtained from the publication by \cite{glowacz2018}. This dataset comprises recordings that have been categorised into two distinct classes: 
\begin{enumerate}
    \item Healthy device, 
    \item Device with a damaged motor, with subcategories:
    \begin{itemize}
        \item Device with damaged gear train (Fault 1),
        \item Device with 5 broken rotor blades (Fault 2),
        \item Device with 10 broken rotor blades (Fault 3), 
        \item Device with a shifted brush (motor off) (Fault 4).
        \end{itemize}
\end{enumerate}
Each class contains 30 one-second-long recordings, which are saved as \texttt{.wav} files with a sampling rate of 4400 Hz. The undamaged device set for the experiment is selected as the first set, while the other sets are used to describe the device with damages. The acquisition of acoustic data from blender and electric impact drill is illustrated in Figures \ref{fig:aqui_1} and \ref{fig:aqui}.

\subsection{Frequency domain analysis}
To conduct the analyses, it was decided to transition to the frequency domain because time series such as audio signals usually have varying lengths. Furthermore, the high sampling frequency means that the analyzed signal, even over a short segment, will contain a large number of samples, which would complicate its analysis. Additionally, in the frequency domain, there is no issue with phase shifts.

For the purposes of the experiments, each signal was divided into equal parts of one second that contains enough information in the FFT. For each signal, the FFT was calculated, and the signal range was limited to the audible frequency range (16Hz-20kHz) due to the characteristics of the microphone used in the case study \cite{porkeba2022functional}.

The FFT transforms (Fig. \ref{fig:FFT}) for the set containing signals of properly functioning devices are presented in Fig. \ref{fig:hsig}. Figure \ref{fig:uhsig} depicts the FFT transforms of signals recorded for faulty devices. The maximum amplitude and the shape of the characteristics in Fig. \ref{fig:FFT} can be identified as the primary differences.

\section{Fault detection results}
\label{sec:FaultDetection}

\begin{figure}[!tb]
    \centering
    
\begin{tikzpicture}[
    scale=1.2,
    shorten >=1pt,->,draw=black!70, node distance=\layersep,
    neuron/.style={circle,fill=black!25,minimum size=20,inner sep=0},
    edge/.style 2 args={pos={(mod(#1+#2,2)+1)*0.33}, font=\tiny},
    distro/.style 2 args={
        edge={#1}{#2}, node contents={}, minimum size=0.6cm, path picture={\draw[double=orange,white,thick,double distance=1pt,shorten >=0pt] plot[variable=\t,domain=-1:1,samples=51] ({\t},{0.2*exp(-100*(\t-0.05*(#1-1))^2 - 3*\t*#2))});}
      },
    weight/.style 2 args={
        edge={#1}{#2}, node contents={\pgfmathparse{0.35*#1-#2*0.15}\pgfmathprintnumber[fixed]{\pgfmathresult}}, fill=white, inner sep=2pt
      }
  ]

  \begin{scope}[xshift=8cm]
    \nn{bayes}
  \end{scope}

  % Draw distros for all Bayesian edges.
  \foreach \i in {1,2,3}
  \foreach \j in {1,...,5}
  \path (X\i-bayes) -- (H\j-bayes) node[distro={\i}{\j}];
  \foreach \i in {1,...,5}
  \path (H\i-bayes) -- (Y-bayes) node[distro={\i}{1}];
\end{tikzpicture}

\caption{The diagram describes the structure of the Bayesian neural network used in the experiments. The developed network takes the successive values of the calculated FFT signal as input arguments, $X1$,...,$X5$, which are passed to five hidden layers, $H1$,...,$H5$, whose activation function is $\tanh$. The output of the network, $Y$, uses the sigmoid function as an activation function. In contrast to traditional neural networks, the weights are represented by distributions}
    \label{fig:bayesian_schema}
\end{figure}
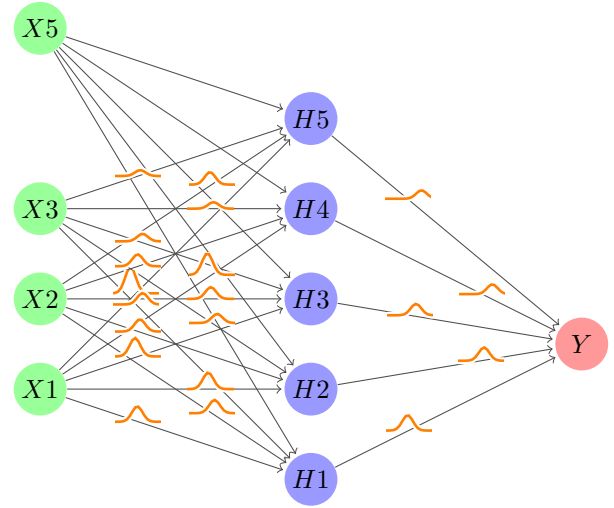

\subsection{Structure of the Bayesian neural network}

For the experiment, BNNs structure was developed, as shown in Fig. \ref{fig:bayesian_schema}. In contrast to traditional neural networks, the weights are represented by distributions. The BNN that was created uses consecutive calculated FFT signal values as input arguments $X1$,...,$X5$, which are passed to five hidden layers, $H1$,...,$H5$. The activation function of these hidden layers is $\tanh$. Prior distribution for weight coefficients for all layers was selected as standard normal distribution. It is justified because of regularization properties of the normal distribution and that signals were reasonably scaled. The output $Y$ of the neural network employs the sigmoid function as its activation function \cite{Kucukelbir2016,mac2003}.

\begin{figure}[H]
\centering
  \includegraphics[width=0.8\columnwidth]{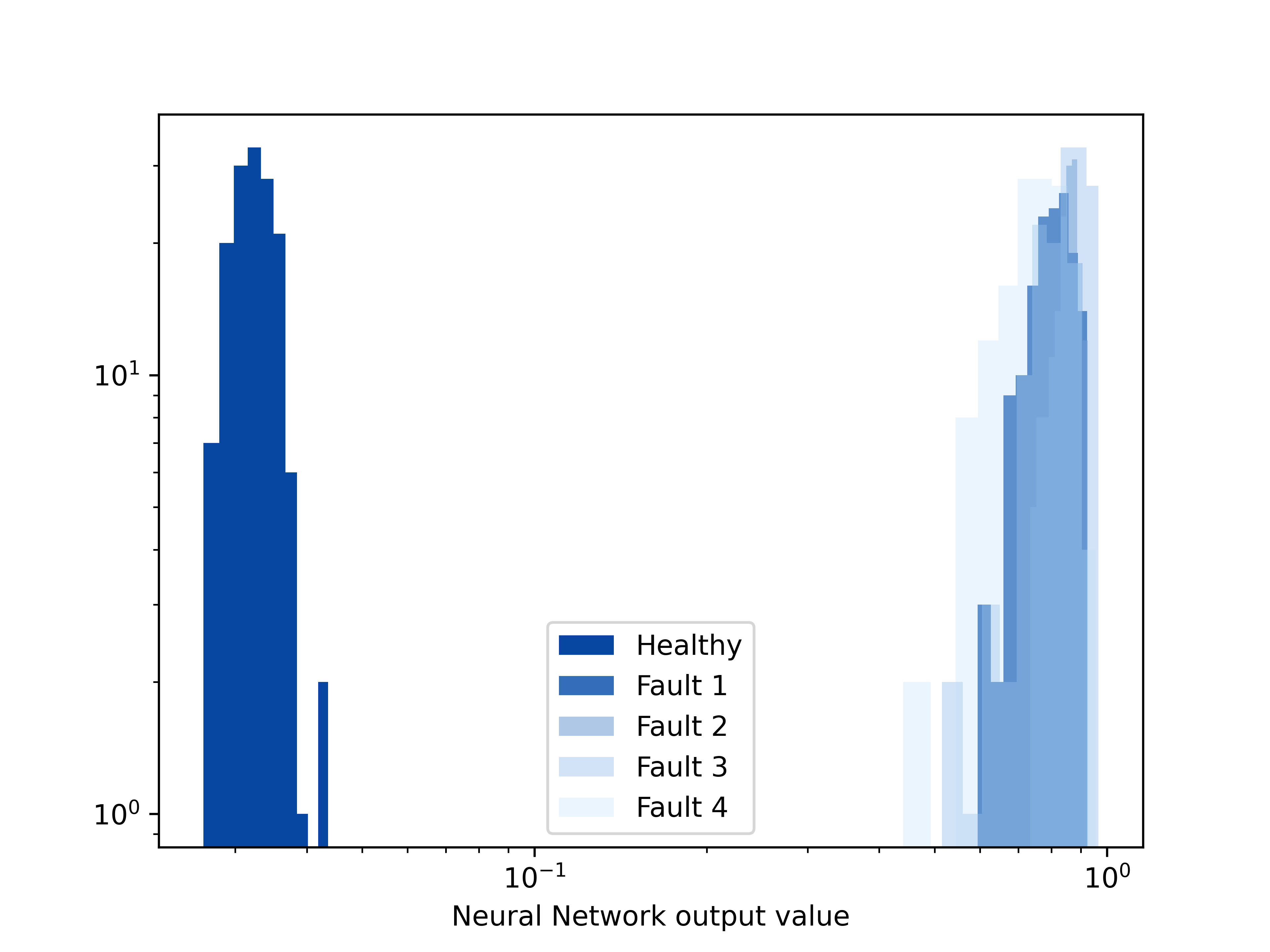}
\caption{Comparison of the distributions of output values of the designed BNN for all signal classes.
\label{fig:result_hist}}
\end{figure}

\begin{figure}[H]
\centering
  \includegraphics[width=0.8\columnwidth]{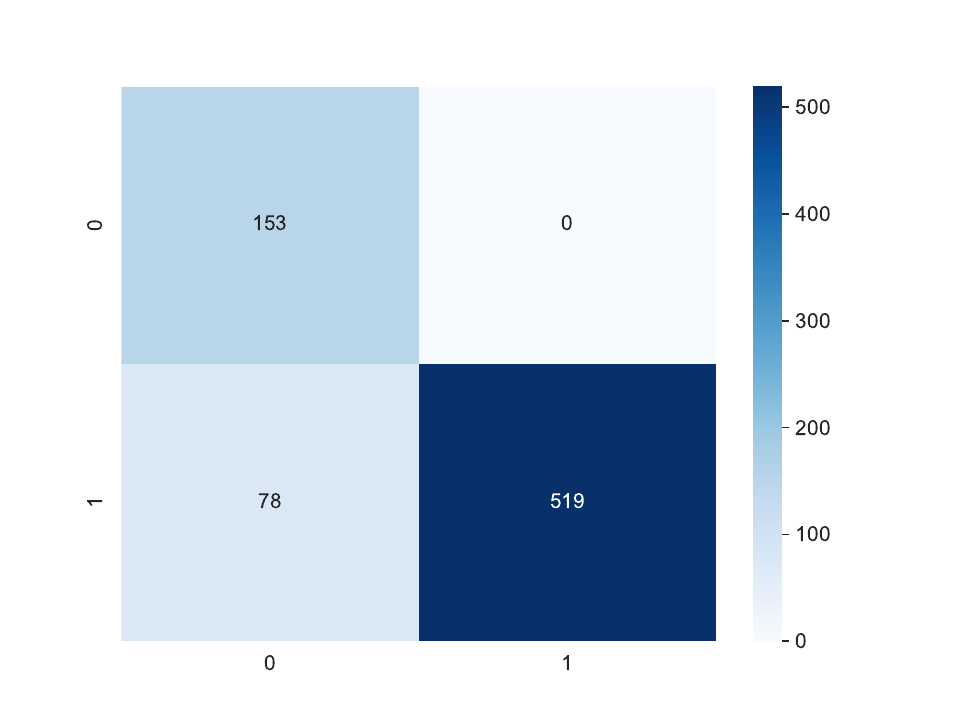}
\caption{Confusion matrix for one of the 100 trials of the developed BNN. The proposed structure flawlessly detects damages and correctly recognizes signals from undamaged devices in almost $70\%$ of cases. This is a desirable property because it accurately informs about the occurrence of a malfunction in the system.
\label{fig:conf_mat}}
\end{figure}

\subsection{Experimental result}

\begin{table}[!tbp]
\centering
\caption{Statistical distribution of predictor values for specific signal classes}
\label{tab:com_class}
\begin{tabular}{@{}ccc@{}}
\toprule
Class name & Predictor mean value & Predictor standard deviation \\ \midrule
Healthy    & 0.046                & 0.211                        \\
Fault 1    & 0.824                & 0.367                        \\
Fault 2    & 0.835                & 0.368                        \\
Fault 3    & 0.866                & 0.324                        \\
Fault 4    & 0.779                & 0.379                        \\ \bottomrule
\end{tabular}%

\end{table}
The implementation and experimentation were carried out using the PyMC 5 library \cite{pymc2023}, which enables construction of Bayesian models with a Python API and fitting these models using Markov Chain Monte Carlo (MCMC) techniques.

Each signal was assigned a class: Signals from a non-damaged device belong to class 0, while the remaining signals belong to class 1. To verify the accuracy of the developed Bayesian network, the dataset was randomly divided into a training set and a test set in the proportion 80/20.

An analysis of the results was then conducted. This process was repeated 100 times, and the average accuracy of the network was approximately $85\%$. An example confusion matrix is presented in Fig. \ref{fig:conf_mat}. The displayed tables demonstrate the precision of the developed BNN. The suggested design detects damages flawlessly (top right) and accurately identifies signals from undamaged devices in nearly $70\%$ of instances. This is a desirable characteristic as it informs about malfunctions in the system with high accuracy. 

The developed classifier effectively distinguishes between properly functioning (Healthy) and malfunctioning devices (Fault 1, ..., 4). However, as indicated by the predictor mean value in Table \ref{tab:com_class}, proposing a classifier capable of differentiating between various types of damage remains an open research question.

\section{Conclusion}
\label{sec:Conclusions}
Bayesian neural networks (BNNs) have been applied to fault detection in various fields, including gas turbine engines, wind turbines, and chemical processes. They detect faults more accurately than traditional neural networks, even for imbalanced or incomplete datasets. In this paper, we focus on fault diagnosis of electric motors using acoustic signals. Such diagnostic signals in time-domain have varying length and are sampled with high frequency, thus necessitating a new signal representation for classification using BNNs. 

We propose an approach based on frequency domain instead of time domain, which allows overcoming these challenges. The Bayesian neural network developed for classification of faults of electric motors of domestic appliances proved 100\% accurate for detecting faulty devices and 70\% accurate for detection of of healthy devices. We can see that Bayesian Neural Network could potentially be useful for diagnostic problems and is well-suited for use in industrial diagnostics systems.

The open problems of this method are chosen to signal description and in the result to provide better performance on this method.  In next step we want start research on the method a functional data fusion to join signals in diagnostic methods. Additionally, there are plans to attempt the development of a classifier capable of differentiating between various classes of damage.

\begin{figure*}[!ht]
     \centering
     \begin{subfigure}{0.4\textwidth}
         \centering
  \includegraphics[width=1\columnwidth]{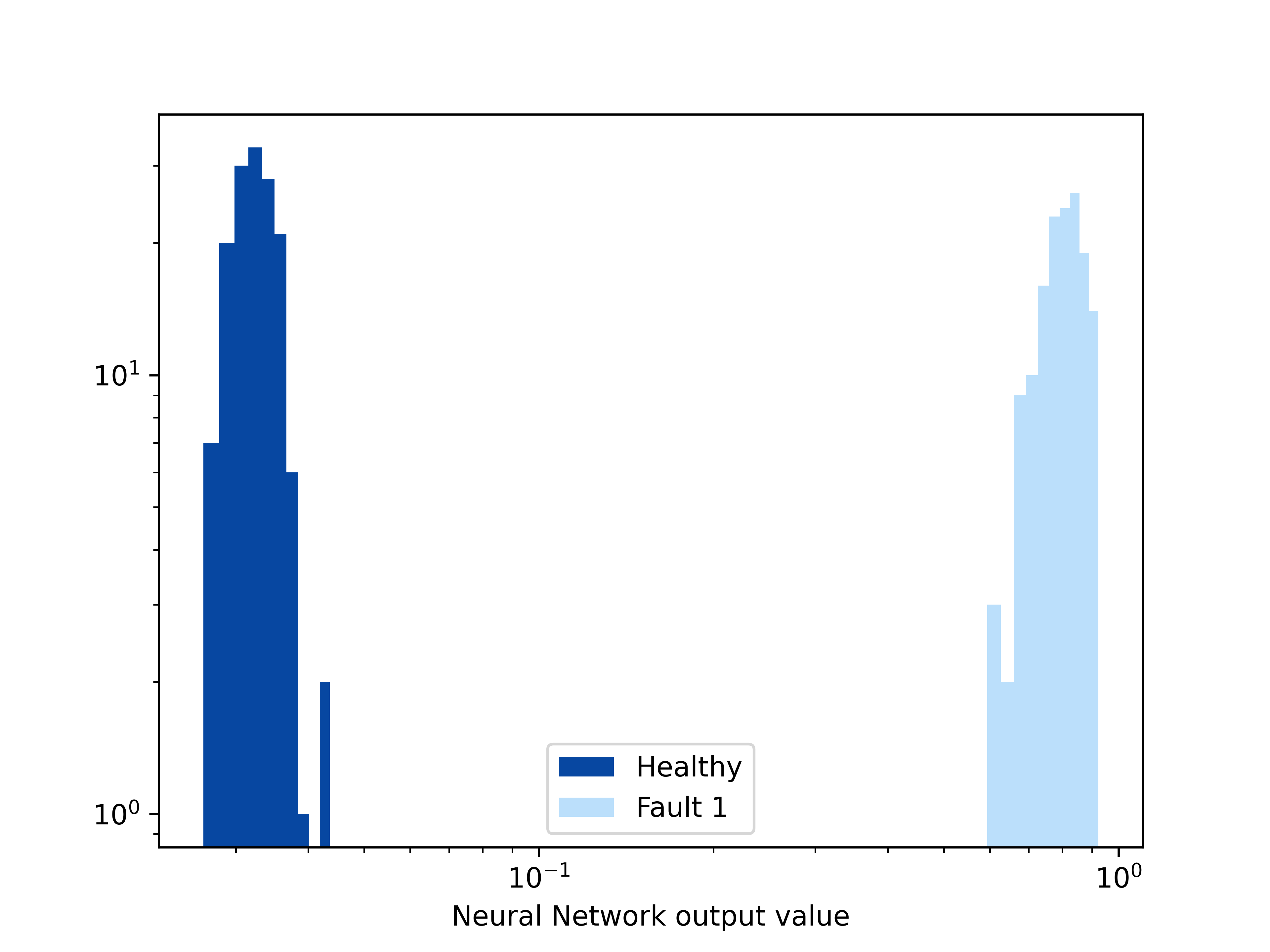}
\caption{Comparison of the distributions of output values of the designed BNN for health and fault 1 signal.\label{fig:result_hist_f1}}
\includegraphics[width=1\columnwidth]{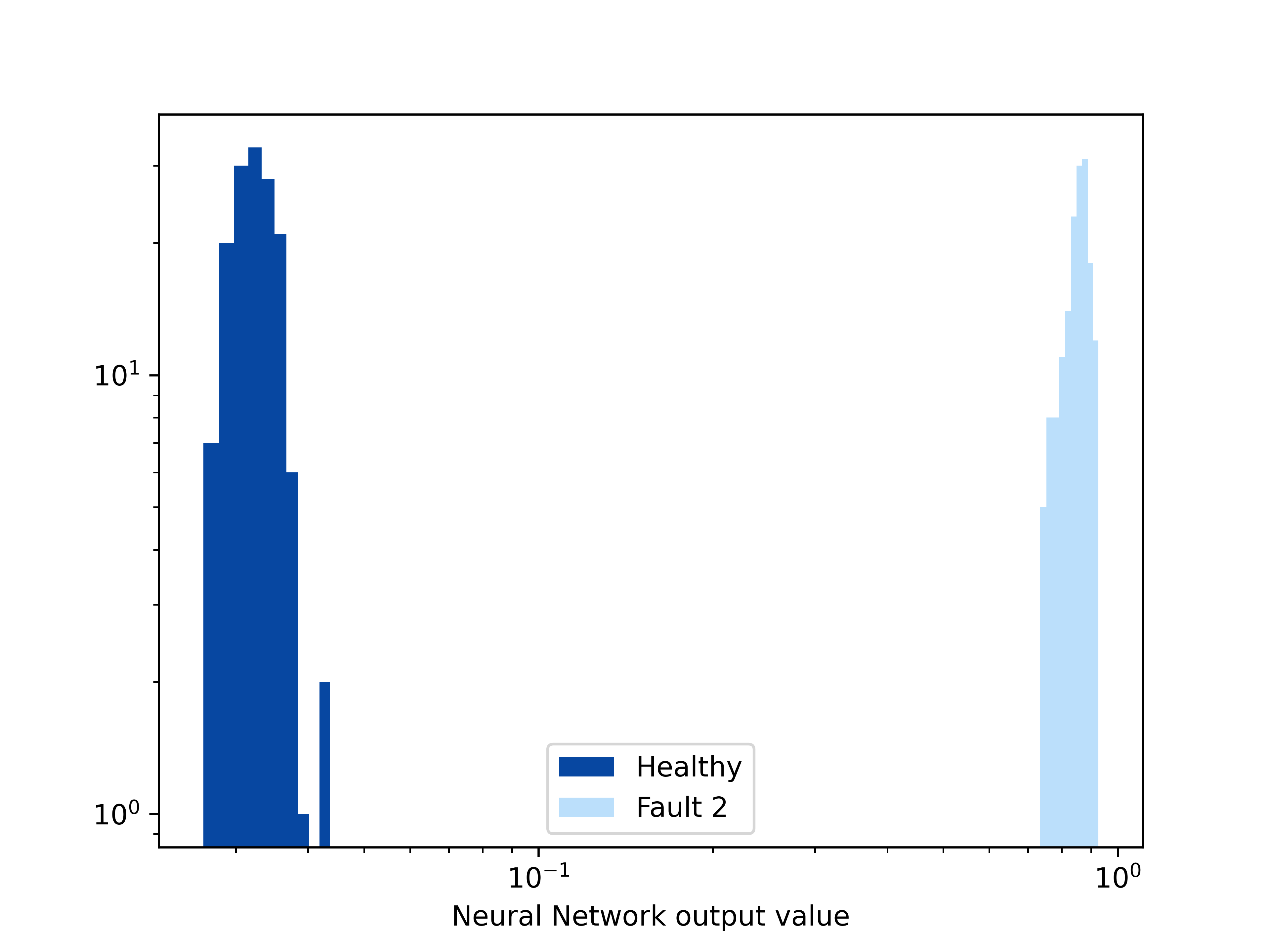}
\caption{Comparison of the distributions of output values of the designed BNN for health and fault 2 signal.\label{fig:result_hist_f2}}
     \end{subfigure}
     \hfill
     \begin{subfigure}{0.4\textwidth}
         \centering
  \includegraphics[width=1\columnwidth]{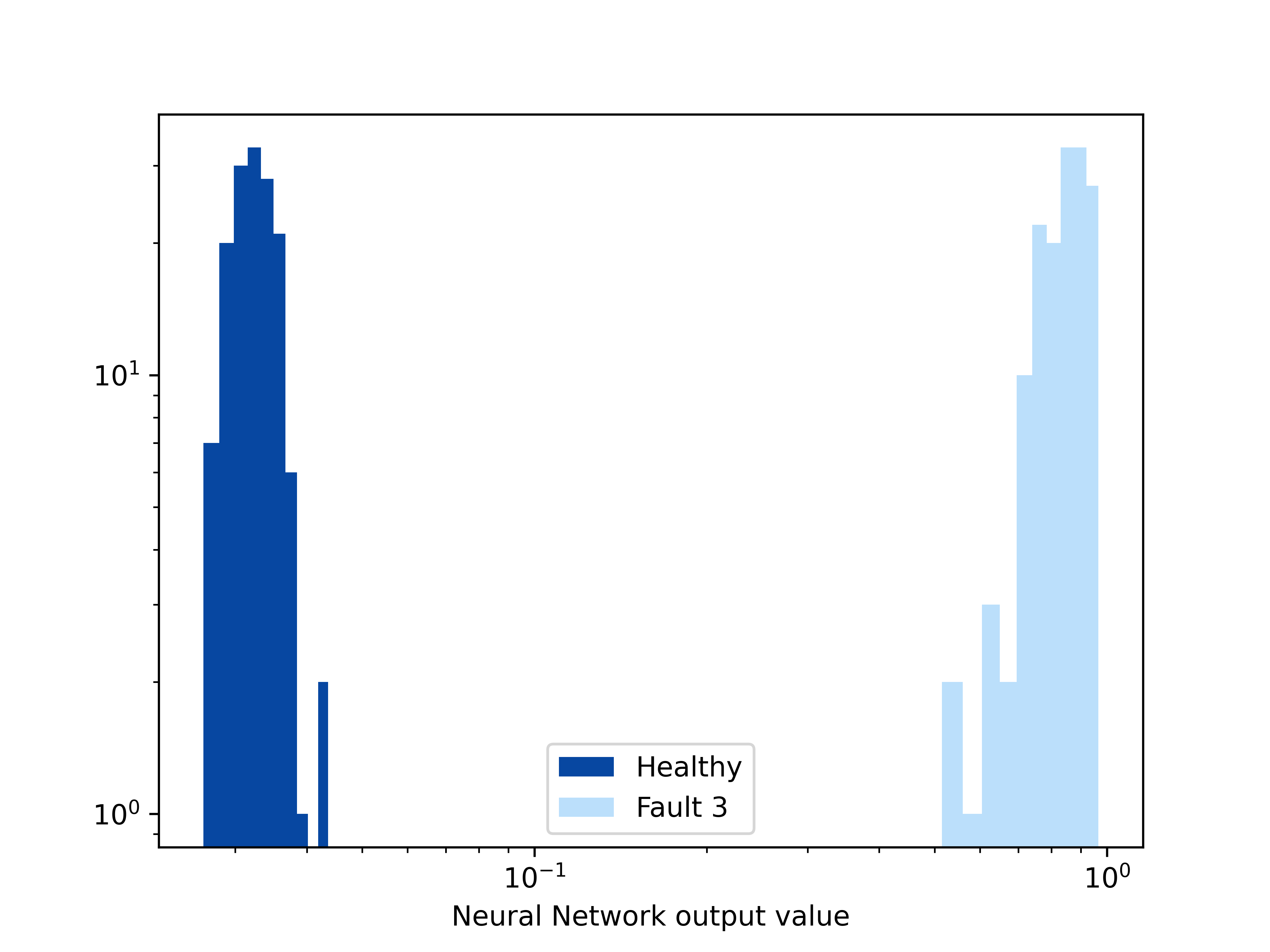}
\caption{Comparison of the distributions of output values of the designed BNN for health and fault 3 signal. \label{fig:result_hist_f3}}
  \includegraphics[width=1\columnwidth]{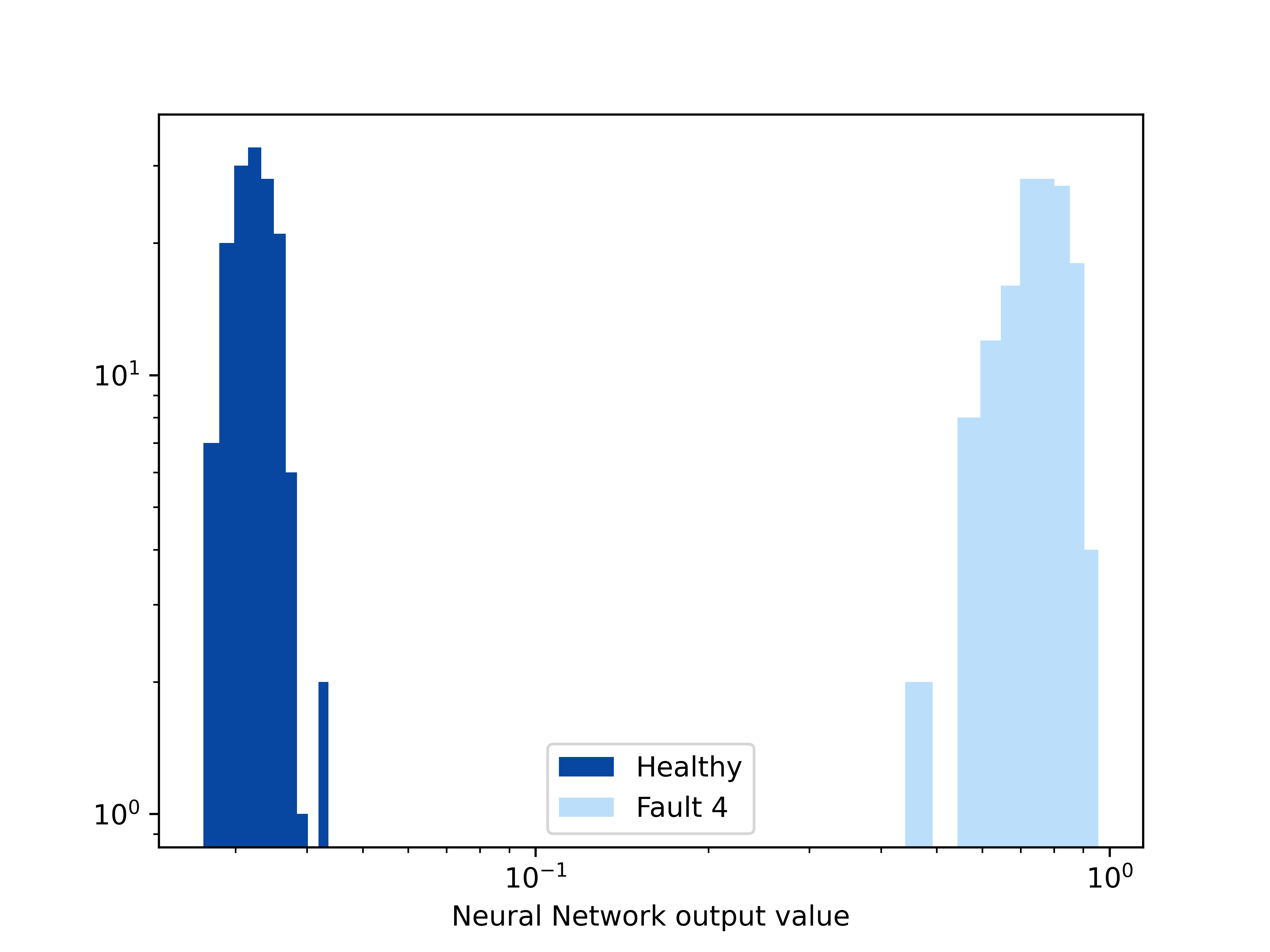}
\caption{Comparison of the distributions of output values of the designed BNN for health and fault 4 signal. \label{fig:result_hist_f4}}
     \end{subfigure}
     \hfill
        \caption{Comparison of the distributions of output values of the designed BNN for a healthy signal and individual classes of faults.}
        \label{fig:com_hist}
\end{figure*}

\bibliographystyle{IEEEtran}
\bibliography{IEEEabrv,bibliography}

\begin{thebibliography}{10}
\providecommand{\url}[1]{#1}
\csname url@samestyle\endcsname
\providecommand{\newblock}{\relax}
\providecommand{\bibinfo}[2]{#2}
\providecommand{\BIBentrySTDinterwordspacing}{\spaceskip=0pt\relax}
\providecommand{\BIBentryALTinterwordstretchfactor}{4}
\providecommand{\BIBentryALTinterwordspacing}{\spaceskip=\fontdimen2\font plus
\BIBentryALTinterwordstretchfactor\fontdimen3\font minus
  \fontdimen4\font\relax}
\providecommand{\BIBforeignlanguage}[2]{{%
\expandafter\ifx\csname l@#1\endcsname\relax
\typeout{** WARNING: IEEEtran.bst: No hyphenation pattern has been}%
\typeout{** loaded for the language `#1'. Using the pattern for}%
\typeout{** the default language instead.}%
\else
\language=\csname l@#1\endcsname
\fi
#2}}
\providecommand{\BIBdecl}{\relax}
\BIBdecl

\bibitem{Tran2023}
M.~Tran, M.~Amer, A.~Dababat, A.~Y. Abdelaziz, H.-J. Dai, M.-K. Liu, and
  M.~Elsisi, ``Robust fault recognition and correction scheme for induction
  motors using an effective iot with deep learning approach,''
  \emph{Measurement}, vol. 207, p. 112398, 2023.

\bibitem{Qiu2023}
S.~Qiu, X.~Cui, Z.~Ping, N.~Shan, Z.~Li, X.~Bao, and X.~Xu, ``Deep learning
  techniques in intelligent fault diagnosis and prognosis for industrial
  systems: A review,'' \emph{Sensors}, vol.~23, no.~3, p. 1305, Jan. 2023.

\bibitem{Tama2022}
B.~A. Tama, M.~Vania, S.~Lee, and S.~Lim, ``Recent advances in the application
  of deep learning for fault diagnosis of rotating machinery using vibration
  signals,'' \emph{Artificial Intelligence Review}, vol.~56, no.~5, pp.
  4667--4709, Oct. 2022.

\bibitem{Zhang2023}
L.~Zhang, Q.~Fan, J.~Lin, Z.~Zhang, X.~Yan, and C.~Li, ``A nearly end-to-end
  deep learning approach to fault diagnosis of wind turbine gearboxes under
  nonstationary conditions,'' \emph{Engineering Applications of Artificial
  Intelligence}, vol. 119, p. 105735, Mar. 2023.

\bibitem{Zhu2023}
Z.~Zhu, Y.~Lei, G.~Qi, Y.~Chai, N.~Mazur, Y.~An, and X.~Huang, ``A review of
  the application of deep learning in intelligent fault diagnosis of rotating
  machinery,'' \emph{Measurement}, vol. 206, p. 112346, Jan. 2023.

\bibitem{mac2003}
D.~J.~C. MacKay, \emph{Information Theory, Inference and Learning
  Algorithms}.\hskip 1em plus 0.5em minus 0.4em\relax UK: Cambridge University
  Press, 2003.

\bibitem{10.5555/3045390.3045502}
Y.~Gal and Z.~Ghahramani, ``Dropout as a {B}ayesian approximation: Representing
  model uncertainty in deep learning,'' in \emph{Proceedings of the 33rd
  International Conference on International Conference on Machine Learning -
  Volume 48}, ser. ICML'16.\hskip 1em plus 0.5em minus 0.4em\relax JMLR.org,
  2016, p. 1050–1059.

\bibitem{Hinton2012}
G.~E. Hinton, N.~Srivastava, A.~Krizhevsky, I.~Sutskever, and R.~R.
  Salakhutdinov, ``Improving neural networks by preventing co-adaptation of
  feature detectors,'' 2012, arxiv preprint, \url{
  https://doi.org/10.48550/arXiv.1207.0580}.

\bibitem{Bakri2017}
A.~E. Bakri, Y.~Berrada, and I.~Boumhidi, ``Bayesian regularized artificial
  neural network for fault detection and isolation in wind turbine,'' in
  \emph{2017 Intelligent Systems and Computer Vision ({ISCV})}.\hskip 1em plus
  0.5em minus 0.4em\relax {IEEE}, apr 2017.

\bibitem{Loboda2015}
I.~Loboda and M.~A.~O. Robles, ``Gas turbine fault diagnosis using
  probabilistic neural networks,'' \emph{International Journal of Turbo
  Jet-Engines}, vol.~32, no.~2, jan 2015.

\bibitem{Ma2020}
X.~Ma, Y.~Yang, L.~Lu, and F.~Liu, ``Chemical process fault detection based on
  bayesian neural networks with dropout,'' \emph{IEEE Access}, 2020.

\bibitem{Rezende2014}
D.~J. Rezende, S.~Mohamed, and D.~Wierstra, ``Stochastic backpropagation and
  approximate inference in deep generative models,'' in \emph{Proceedings of
  the 31st International Conference on Machine Learning}, ser. Proceedings of
  Machine Learning Research, E.~P. Xing and T.~Jebara, Eds., vol.~32,
  no.~2.\hskip 1em plus 0.5em minus 0.4em\relax Bejing, China: PMLR, 22--24 Jun
  2014, pp. 1278--1286.

\bibitem{glowacz2018}
A.~Glowacz, ``Recognition of acoustic signals of commutator motors,''
  \emph{Applied Sciences}, vol.~8, no.~12, p. 2630, dec 2018.

\bibitem{porkeba2022functional}
J.~Por{\k{e}}ba and J.~Baranowski, ``Functional logistic regression for motor
  fault classification using acoustic data in frequency domain,''
  \emph{Energies}, vol.~15, no.~15, p. 5535, 2022.

\bibitem{Kucukelbir2016}
A.~Kucukelbir, D.~Tran, R.~Ranganath, A.~Gelman, and D.~M. Blei, ``Automatic
  differentiation variational inference,'' \emph{Journal of machine learning
  research}, vol.~18, no.~14, pp. 1--45, 2017.

\bibitem{pymc2023}
O.~Abril-Pla, V.~Andreani, C.~Carroll, L.~Dong, C.~J. Fonnesbeck, M.~Kochurov,
  R.~Kumar, J.~Lao, C.~C. Luhmann, O.~A. Martin, M.~Osthege, R.~Vieira,
  T.~Wiecki, and R.~Zinkov, ``{PyMC}: A modern and comprehensive probabilistic
  programming framework in {Python},'' \emph{{PeerJ} Computer Science}, vol.~9,
  p. e1516, 2023.

\end{thebibliography}

\end{document}